\documentclass[aps,pre,twocolumn,superscriptaddress,tight,showpacs]{revtex4-2}

\usepackage{amssymb}
\usepackage{color}
\usepackage{graphicx}
\usepackage{amsmath}
\usepackage{mathrsfs}
\usepackage{times}
\usepackage{subeqnarray}
\usepackage{cases}
\usepackage{bm}
\usepackage{float}
\usepackage{import}

\begin{document}

\title{Breathing cluster in complex neuron-astrocyte networks}

\author{Ya Wang}
\affiliation{School of Physics and Information Technology, Shaanxi Normal University, Xi'an 710062, China}
\author{Liang Wang}
\affiliation{School of Physics and Information Technology, Shaanxi Normal University, Xi'an 710062, China}
\author{Huawei Fan}
\affiliation{School of Science, Xi’an University of Posts and Telecommunications, Xi'an 710121, China}
\author{Jun Ma}
\affiliation{Department of Physics, Lanzhou University of Technology, Lanzhou 730050, China}
\author{Hui Cao}
\email[Email address: ]{caohui@snnu.edu.cn}
\affiliation{School of Physics and Information Technology, Shaanxi Normal University, Xi'an 710062, China}
\author{Xingang Wang}
\email[Email address: ]{wangxg@snnu.edu.cn}
\affiliation{School of Physics and Information Technology, Shaanxi Normal University, Xi'an 710062, China}

\begin{abstract}
Brain activities are featured by spatially distributed neural clusters of coherent firings and a spontaneous switching of the clusters between the synchrony and asynchrony states. Evidences from {\it in vivo} experiments suggest that astrocytes, a type of glial cell regarded previously as providing only structural and metabolic supports to neurons, participate actively in brain functions and play a crucial role in regulating the neural firing activities, yet the mechanism remains unknown. Introducing astrocyte as a reservoir of the glutamate released from neuron synapses, here we propose the model of complex neuron-astrocyte network and employ it to explore the roles of astrocyte in regulating the synchronization behaviors of networked neurons. It is found that a fraction of neurons on the network can be synchronized as a cluster, while the remaining neurons are kept as desynchronized. Moreover, during the course of network evolution, the cluster is switching between the synchrony and asynchrony states intermittently, henceforth the phenomenon of ``breathing cluster". By the method of symmetry-based analysis, we conduct a theoretical investigation on the stability of the cluster and the mechanism generating the breathing activities. It is revealed that the contents of the cluster are determined by the network symmetry and the breathing activities are due to the interplay between the neural network and the astrocyte. The breathing phenomenon is demonstrated in network models of different structures and neural dynamics. The studies give insights into the cellular mechanism of astrocytes in regulating neural activities, and shed lights onto the spontaneous state switching of the neocortex. 

\end{abstract}
\maketitle

\section{INTRODUCTION}

Theoretical studies and experimental evidences suggest that synchronized firing activities of neurons within and across different brain regions are a fundamental property of cortical and subcortical networks and underpin the functions of a variety of cognitive processes~\cite{Singer1999,BrainWebSyn,Ward2003,Rev:Uhlhaas2006,Hipp:Neuron2011,Rev:JaJun2015,Rev:LZH2022}. Whereas numerous neural models have been proposed for exploring the mechanism of the firing patterns~\cite{Rev:JaJun2015,Rev:LZH2022}, the models are dominantly based on neurons, though neurons are outnumbered by glial cells in the brain~\cite{Glia:Barres,Glia:VA}. In particular, stimulated by the blooming of network science and the discoveries of the small-world and scale-free features in brain networks~\cite{SW:1998,BA:1999,Brain:SW,Brain:SFN,NetworkSynBrain,NetNeurosci2017}, intensive theoretical studies have been conducted to explore the interplay between neural network structure and collective firing activities, in which important insights have been gained into the organization and functionality of the brain. 

In neuroscience, a prevalent opinion is that the cognitive functions of the brain rely on solely the collective firing of the neural cells, which are embraced by glial cells and the roles of the glial cells are providing only the structural and metabolic supports to neurons~\cite{Book:KER,Book:glia}. This opinion, however, has been changed largely in recent years, as accumulating experimental evidences suggest that glial cells are active partners of neurons in both signal processing and information integration, and play a vital role in maintaining the normal functioning of the brain~\cite{GlialCell:HPG,GlailCell:RDF}. A typical example is astrocytes (also known as astroglia), which are the most abundant glial cells in the brain and participate actively in many brain functions through modulating the synaptic transmission and plasticity of the cortical synapses~\cite{Astro:Science1990,Astro:Nature1994,Astro:HRP2001,Astro:GP2007,Astro:OY2013,Rev:astrocyte2021}. In specific, experimental studies show that by releasing glutamate to the synaptic clefs and the extracellular space, astrocytes modulate effectively the synchrony of the neural firings, e.g., triggering the transient synchronization a group of neurons by stimulating a single astrocyte~\cite{GliaExpSyn:TF2004,GliaExpSyn:MCA2004,UpdtateAstro2011,Poskanzer:Switching2016}. Despite the experimental advances, the molecular and cellular mechanisms by which astrocytes regulate neuron synchronization remain poorly understood, calling for the development of sophisticated network models capturing the ``cross-talk" between neurons and astrocytes~\cite{Rev:astrocyte2021}. 

Models have been proposed in the literature for exploring the interplay between neurons and astrocytes, yet are restricted to small-size systems and the studies are focusing on the special state of global synchronization~\cite{SynModel:SN2003,SynModel:PA2009,SynModel:AP2009,SynModel:DEP2009,SynModel:MJ2013,SynModel:Amiri2011,SynModel:Amiri2012,SynModel:Amiri2013,TriSynap:19999,SynModel:SYM2020,SynModel:SY2021}. In terms of system size, most of the studies are based on minimal networks consisting of a handful of elements~\cite{SynModel:SN2003,SynModel:PA2009,SynModel:AP2009,SynModel:MJ2013,SynModel:SYM2020}, e.g., a tripartite network containing one astrocyte and two neurons~\cite{SynModel:Amiri2011,SynModel:Amiri2013,TriSynap:19999}, which are clearly different from the cortical networks in size (consisting of a large number of neurons and astrocytes) and structure (represented by complex networks)~\cite{GliaExpSyn:TF2004,GliaExpSyn:MCA2004,UpdtateAstro2011,Poskanzer:Switching2016}. Though the modulating effects of astrocytes on neural synchrony have been demonstrated in the minimal network models, the generalization of the findings to complex networks is yet to be checked, especially when considering the significant impacts of network structure on synchronization dynamics. In terms of synchronization, whereas complex neuron-astrocyte networks have been adopted in some studies~\cite{SynModel:Amiri2011,SynModel:Amiri2013,TriSynap:19999}, the synchrony states investigated are stationary in time and uniform in space, e.g., the state of global synchronization, which are also different from the experimental observations that brain cognitive functions are supported by the transient synchrony of clustered neurons~\cite{TranSyn:MS:2001,TranSyn:RC2003,BrainOscillation}. Different from the previous modeling studies, here we focus on the synchronization of neural clusters on complex neuron-astrocyte networks, report the new phenomenon of cluster breathing and analyze the mechanism generating the breathing. 

Our present work is also inspired by the recent studies on cluster synchronization in complex networks~\cite{SynSymtry:Russo2011,CS:VN2013,Pecora2014,LWJ-1,FS:2016,LWJ-2,YC:2017,CS:WYF2019}. In exploring the synchronization dynamics of coupled oscillators, an interesting phenomenon is that in some circumstances the oscillators can be self-organized into different clusters, with the motions of the oscillators within each cluster being highly correlated but not for oscillators from different clusters. This phenomenon, known as cluster (partial or group) synchronization~\cite{CS:Hasler,CS:YZ,CS:AP,CS:CRSW}, has important implications to the functionality and operation of many natural and man-made systems, and has been broadly interested and extensively investigated in the fields of nonlinear science and complex systems in the past decades~\cite{CS:RevADM2016}. Recently, with the developments of computational group algorithms and sophisticated mathematical techniques, significant progress has been made in the study of cluster synchronization in large-size complex networks, in which the important roles of network symmetries have been revealed~\cite{SynSymtry:Russo2011,CS:VN2013,Pecora2014,LWJ-1,FS:2016,LWJ-2,YC:2017,CS:WYF2019}. In cluster synchronization, a special case is that a subset of oscillators are completely synchronized and form a cluster, while the other oscillators on the network are remaining as desynchronized. This phenomenon, termed isolated desynchronization (IDS) in Ref.~\cite{Pecora2014} and independently synchronizable cluster (ISC) in Ref.~\cite{YC:2017}, is reminiscent of the UP states observed in the cortical cortex, in which a group of neurons fire action potentials coordinately for up to hundreds of milliseconds~\cite{TranSyn:MS:2001,TranSyn:RC2003,UpdtateAstro2011}. However, clusters in the IDS and ICS states are stable and stationary, in the sense that oscillators within the cluster are synchronized throughout the process of system evolution. These features make the IDS and ICS states different from the UP states, as in the latter the neural clusters are switching intermittently between the synchrony and asynchrony states at a slow frequency~\cite{Poskanzer:Switching2016,UpdtateAstro2011}. From the point of view of network modeling, a question of theoretical interest is how to make the cluster switch spontaneously between the synchrony and asynchrony states, especially for a large-size complex network. We are going to demonstrate and argue that this question can be addressed by incorporating astrocytes into the conventional model of neural networks, and the interplay between astrocytes and neurons leads to naturally a slow switching of the cluster between the synchrony and asynchrony states, thereby reproducing the phenomenon of state switching observed in experiments.

The rest of the paper is organized as follows. In Sec.~II, we will first describe the interacting mechanism between neurons and astrocytes, and then propose the model of complex neuron-astrocyte network. In Sec.~III, we will study by numerical simulations the firing activities of the neurons, report the phenomenon of cluster breathing and investigate the impacts of the astrocyte parameters on cluster breathing. In Sec.~IV, we will conduct a theoretical analysis on the stability of the synchronization cluster, and explore the mechanism underlying cluster breathing. In Sec.~V, we will generalize the findings to other network models, including complex networks of different sizes, different neural dynamics, and different coupling functions. We wrap up the paper with discussions and conclusions, which will be given in Sec.~VI.

\section{Complex neuron-astrocyte network}

\begin{figure*}[tbp]
\begin{center}
\includegraphics[width=0.7\linewidth]{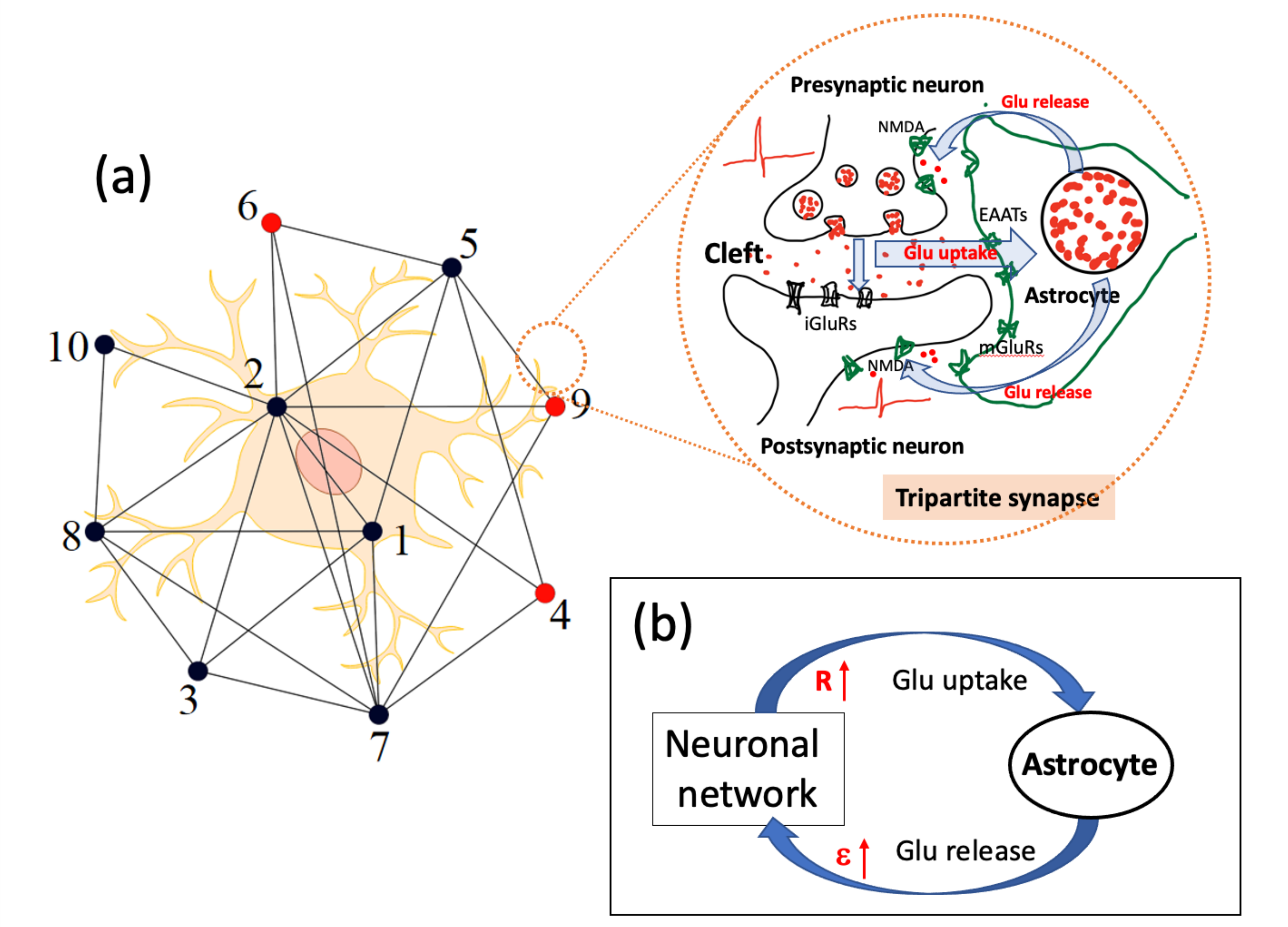}
\caption{(Color online) Schematic of the complex neuron-astrocyte network. (a) The network structure. The network consists of $N=10$ randomly connected neurons and a single astrocyte (the background cell). Neurons $4$, $6$ and $9$ are symmetric. Inset: schematic of a tripartite synapse made up of presynaptic and postsynaptic terminals and astrocyte process. About $20\%$ of the glutamate released from the presynaptic terminal to the synaptic cleft is bonded to the postsynaptic terminal, and the remaining ($\sim 80\%$) is taken up by the astrocyte. When activated by the Ca$^{2+}$ signals, glutamate will be released from the astrocyte to the extracellular space, which promotes the coherent firing of the adjacent neurons. (b) Schematic of the interplay between neural firing activities and astrocytic dynamics. Please see the context for more details.}
\label{fig1}
\end{center}
\end{figure*}

Whereas abundant network models have been proposed for exploring the neural activities of the cortical cortex at different spatial scales, few models consider neuron-astrocyte interactions~\cite{Rev:astrocyte2021}. Astrocytes tile the cortex with nearly complete coverage and are connected into an extensive complex network through gap junctions~\cite{Astro:EAB2002,AstroNet:HMM2007}. As such, a more accurate model of the brain cortex should include at least two interacting complex networks, one for neurons and the other one for astrocytes. However, the territorial feature of astrocytes makes it reasonable to represent the cortical network by a simplified model. Anatomical data suggest that a single astrocyte can contact up to tens of thousands of neural synapses~\cite{Astro:EAB2002}, thereby capable of influencing the activities of a large ensemble of neurons. In the meantime, astrocytic processes are short, and each astrocyte influences mainly its neighboring neurons. Therefore, focusing on only a small cortical area (e.g., a column), the physiological activities of the small area can be modeled by a hybrid network comprising an ensemble of neurons and a single astrocyte, with neurons being connected through synapses while all neurons are interacting with the astrocyte. A schematic of the neuron-astrocyte network is plotted in Fig.~\ref{fig1}(a), which comprises a random network of $N=10$ neurons and one astrocyte. 

Having fixed the structure of the neuron-astrocyte network, we proceed to characterize the interactions between neurons and astrocyte. Astrocyte participates in neural activities mainly by regulating glutamate concentration in the extracellular space, including the processes of glutamate uptake and release~\cite{Astro:Uptake2008,Astro:Release2009}. When a presynaptic neuron is fired, it releases glutamate into the synaptic clefts. A small percentage of the glutamate (about 20$\%$ in the retina and cerebellar cortex and, probably, a lesser percentage in the hippocampus) is taken up by postsynaptic neural receptors such as iGluRs~\cite{AstroGlut:Bergles,AstroGlut:Grant,AstroGlut:Otis}, while the majority of the glutamate (80$\%$ or more) diffuses out of the synaptic clefts and is cleared from the extracellular space by astrocytic transporters such as EAAT-1 and EAAT-2~\cite{AstroGlut:Benediktsson,AstroGlut:Zhou}. Inside the astrocyte, a fraction of the glutamate will be metabolized to tricarboxylic (TCA) for synthesizing ATP, while most of the glutamate is stored in cellular vesicles~\cite{Metab:Sonnewald,Metab:McKenna}. Glutamate update is crucial for maintaining the normal functioning of the central nervous system, as excess glutamate may lead to neural hyperexcitation and subsequent neural death, in a process known as ``glutamate excitotoxicity"~\cite{Excitotoxicity}. The glutamate stored in astrocyte will be released to the extracellular space when the concentration of Ca$^{2+}$ exceeds a certain threshold, namely the process of glutamate release~\cite{Astro:Release2009}. Besides being taken up by astrocyte, some glutamate released by the presynaptic neurons is bounded to the metabotropic glutamate receptors (mGluRs) on the astrocytic membrane. The activation of mGluRs triggers the release of the second messenger inositol (1,4,5)-trisphosphate (IP$_3$), followed by the release of IP$_{3}$-dependent Ca$^{2+}$ from the endoplasmic reticulum (ER) and Ca$^{2+}$-dependent Ca$^{2+}$ from ER. Once the intracellular Ca$^{2+}$ concentration exceeds a certain threshold, it will trigger the release of the stored glutamate into the extracellular space. The released glutamate not only activates the postsynaptic N-methyl-D-aspartate (NMDA) receptors of the adjacent excitatory neurons (which generates an additional inward current and promotes the firing of the postsynaptic neuron)~\cite{GliaExpSyn:TF2004,GliaExpSyn:MCA2004,Poskanzer:Switching2016}, but also activates the mGluRs and NMDA receptors of the presynaptic neurons (which promotes the release of glutamate from the presynaptic neurons to the clefs and thereby enhances the neural excitability)~\cite{Astro:GP2007}. Both pathways increase the strength of synaptic interaction and promote the coherent firing of the adjacent neurons~\cite{Astro:Jourdain}. The uptake and release processes thus give the following picture of neuron-astrocyte interactions: the coherent firing of the neurons increases the glutamate concentration in astrocyte, and the glutamate released from the astrocyte strengthens the synapses and subsequently the synchrony of neural firings. The mutual interaction between neurons and astrocyte is schematically plotted in Fig.~\ref{fig1}(b).

We concrete the neuron-astrocyte network model by setting specific dynamics for neurons and astrocyte. As the territorial domain of each astrocyte covers a large number of densely connected neurons, we model the neural network by a dense network of randomly coupled neuron oscillators. Specifically, we adopt the three-dimensional Hindmarsh-Rose (HR) oscillator to describe the firing dynamics of the isolated neurons~\cite{HR:1984}, and couple an ensemble of HR oscillators by the chemical synapses on a random complex network. (The results to be reported are general and independent of the neural dynamics and network structure, as will be discussed in Sec. V.)  The dynamics of the HR oscillators are governed by equations
\begin{eqnarray}
\begin{cases}
\dot{x}_i=y_i + 3x_i^{2}-x_i^{3} - z_i + I_s + I^i_{c}+D\xi_i(t),\\
\dot{y}_i=1 - 5x_i^{2} - y_i,   \\
\dot{z}_i=\gamma(4 x_i + 6.4 - z_i),\\
\end{cases}
\label{HR}
\end{eqnarray}
with $i=1,\ldots,N$ being the neuron (oscillators) index and $N$ denoting the network size. The variables $x_i$, $y_i$ and $z_i$ represent, respectively, the membrane potential, the transport rates of the fast and slow ion channels of the $i$th neuron. $I_s$ denotes the intensity of the external stimulating current, and $\gamma$ is the parameter characterizing the rate of the slow ion channels. In our studies, we set $\gamma=6\times 10^{-3}$ and $I_s=3.2$ for all neurons, by which the isolated neuron presents the chaotic bursting activity~\cite{HR:1984}. In Eq.~(\ref{HR}), $\xi(t)$ represents the noise perturbations, which is an independent and identically distributed random variable within the range $(-1,1)$. $D$ denotes the noise amplitude. $I^i_c$ is the coupling signal that neuron $i$ receives from its neighboring neurons on the network, which reads   
\begin{equation}
I^i_c=\varepsilon(V_{r}-x_{i})\sum^{N}_{j=1}a_{ij}h(x_{j}),\\
\label{coupling}
\end{equation}
with $h(x_j)=[1 + e^{-\lambda(x_{j}-\alpha)}]^{-1}$ being the coupling function of chemical synapse and $\varepsilon$ denoting the coupling strength. $V_r$ is the reversal potential of the chemical synapse. $\alpha$ and $\lambda$ are parameters characterizing the synaptic function. Without the loss of generality, we set the parameters $(V_r,\lambda,\alpha)=(2,7.5,-0.25)$. As $V_r>x_i$, the synapses are excitatory. (The case of inhibitory synapses will be discussed in Sec.~V.) The coupling relationship of the neurons is captured by the adjacency matrix ${\bm{A}=\{a_{ij}\}}_{N\times N}$: $a_{ij}=a_{ji}=1$ if neurons $i$ and $j$ are connected by a synapse, and $a_{ij}=a_{ji}= 0$ otherwise.  

Astrocyte participates in neural activities by modulating the synaptic strength, $\varepsilon$, in Eq.~(\ref{coupling}). While sophisticated nonlinear models have been proposed in the literature for describing the dynamics of astrocyte~\cite{AstroModel:Young1992,AstroModel:LYX1994,AstroModel:UG:2006}, the models are inherently oscillatory. That is, once the astrocyte is evocated, its variables will be oscillating with time with large amplitudes. To highlight the fact that cluster breathing is an intrinsic feature of the neuron-astrocyte network (but not induced by astrocytic oscillations), here we adopt, intentionally, a linear model for the astrocytic dynamics. Meanwhile, as astrocyte regulates neural activities mainly through the processes of glutamate uptake and release, and also for the reason that glutamate concentration in the extracellular space is positively correlated to the strength of neural synapses~\cite{Astro:Jourdain}, we model the astrocyte as a reservoir of glutamate and assume a linear dependence between the synaptic strength and the glutamate concentration. With these concerns, we propose to model the dynamics of astrocyte by the equation
\begin{equation}
\dot{\varepsilon}=-a\varepsilon+bR(t-\tau)+c,
\label{glu}
\end{equation}    
where $a$ represents the attenuation coefficient (due to glutamate diffusions and metabolic consumptions), $b$ is the generic coupling strength characterizing the influence of the neural activities on astrocytic dynamics, and $c$ is the constant input (for maintaining the basal glutamate concentration). The influence of the neural activities on the astrocytic dynamics is realized through the synchronization order parameter
\begin{equation}
\label{orderparameter}
R(t)=\frac{1}{N}\vert\sum^{N}_{j=1}e^{i\theta_{j}(t)}\vert,\\ \nonumber
\end{equation}
where $\theta_{j}(t)$ is the instant phase of the $j$th neuron defined as
\begin{equation}\label{phase}
\theta_{j}(t)=2\pi\frac{t-t_{j,k}}{t_{j,k+1}-t_{j,k}},\\ \nonumber
\end{equation}
with $t_{j,k}\leq t \leq t_{j,k+1}$, and $t_{j,k}$ and $t_{j,k+1}$ are the time instants of the $k$th and $(k+1)$th firing spikes of neuron $j$~\cite{HR:Phase}. We have $R\in (0,1)$, with larger $R$ representing stronger coherence of the neural firings. Considering the limited speed of glutamate transmission in the extracellular space, we introduce a time delay coefficient, $\tau$, to the synchronization order parameter in Eq.~(\ref{glu}). To be more specific, we make the growing rate of glutamate concentration at time $t$ to be dependent of $R(t-\tau)$. The astrocytic model described by Eq.~(\ref{glu}), though of simplified dynamics as compared with the sophisticated models proposed in the literature~\cite{AstroModel:Young1992,AstroModel:LYX1994,AstroModel:UG:2006}, captures the essential features of neuron-astrocyte interactions and, as to be demonstrated below, is capable of inducing a spontaneous switching of the neural network between different synchronization states.

\section{Breathing synchronization cluster}

\begin{figure*}[tbp]
\begin{center}
\includegraphics[width=0.65\linewidth]{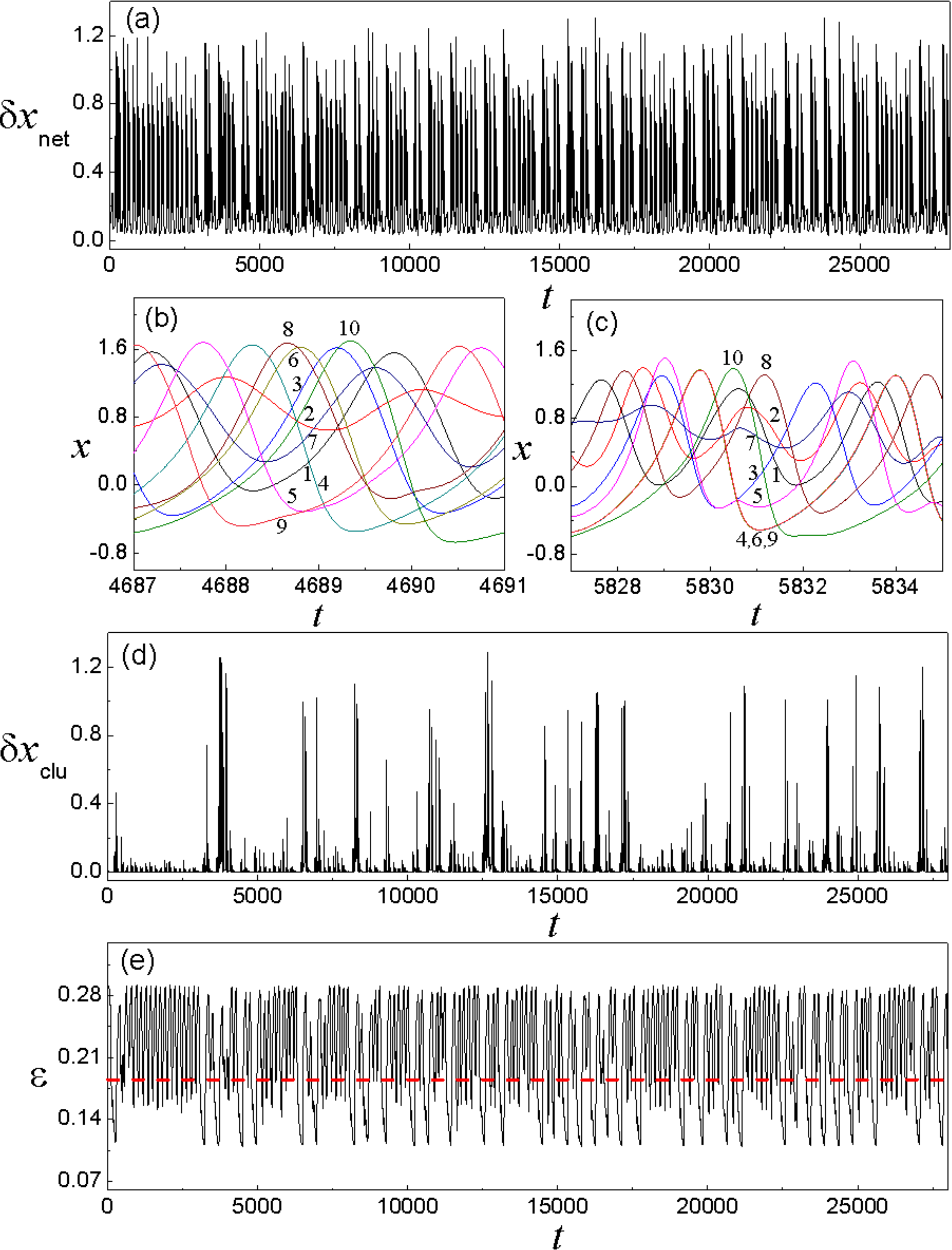}
\caption{(Color online) Breathing of synchronization cluster in complex neuron-astrocyte network. The parameters of the astrocyte are $(a,b,c,\tau)=(3\times 10^{-2}, 8\times 10^{-3}, 1\times 10^{-3}, 200)$. (a) Time evolution of the global network synchronization error $\delta x_{net}$. (b) Typical episode of network desynchronization. (c) Typical episode of cluster synchronization. The trajectories of neurons $4$, $6$ and $9$, are overlapped in (c). (d) Time evolution of the cluster-based synchronization error $\delta x_{clu}$. (e) Time evolution of the coupling strength (the glutamate concentration in astrocyte) $\varepsilon(t)$. Red-dashed line denotes the critical coupling $\varepsilon\approx 0.18$ predicted by the theory for synchronizing the cluster.} 
\label{fig2}
\end{center}
\end{figure*}

We start by investigating numerically the dynamical behaviors of the neuron-astrocyte network constructed above. In simulations, the initial conditions of the neurons and the astrocyte are chosen randomly within the range $(-1,1)$, the initial coupling strength is chosen randomly from the interval $(0,1)$, the noise amplitude is set as $D=0.01$, and the network dynamics, i.e., Eqs.~(\ref{HR}-\ref{glu}), is simulated by the $4$th-order Runge-Kutta algorithm with the time step $\delta t = 0.01$. As the illustration, we set the parameters of the astrocyte as $(a,b,c,\tau)=(3\times 10^{-2}, 8\times 10^{-3}, 1\times 10^{-3}, 200)$ (the impacts of the parameters on the network dynamics will be discussed later). As our interest is focusing on the coherent firing of the neurons, we check first the time evolution of the synchronization degree of the whole neural network. Considering the fact that both firing amplitude and time play important roles in neural computations~\cite{Book:KER}, we characterize the synchronization degree of the neural network by the error $\delta x_{net}(t)=\sum_{i}|x_i(t)-\bar{x}(t)|/N$, with $\bar{x}(t)=\sum_i x_i(t)/N$ the instant membrane potential averaged over all neurons. In general, the smaller the value of $\delta x_{net}$, the stronger the correlation between the neurons and the larger the network synchronization degree. If as time increases the value of $\delta x_{net}$ approaches $0$ gradually, we say that the neural network is synchronizable. Otherwise, if $\delta x_{net}$ keeps at large values after a long-time running, the network is regarded as non-synchronizable. The time evolution of $\delta x_{net}$ is plotted in Fig.~\ref{fig2}(a). We see that $\delta x_{net}$ is oscillating wildly with large amplitudes throughout the process. Clearly, the neural network is non-synchronizable. [Please note that at some time instants $\delta x_{net}$ is close to $0$ ($\sim 0.1$), which is due to the bursting nature of the neural dynamics~\cite{HR:1984}, instead of the occurrence of neuron synchronization.]  

Recent studies of network synchronization show that, even though the whole network is non-synchronizable, a subset of the nodes on the network could be synchronized and form a cluster~\cite{SynSymtry:Russo2011,CS:VN2013,Pecora2014,LWJ-1,FS:2016,LWJ-2,YC:2017,CS:WYF2019}. Inspired by this, we move on to check the firing activities of the network at the neural level by tracing the time evolution of the neural membrane potentials, $x_i(t)$ with $i=1,\ldots,N$. Shown in Figs.~\ref{fig2}(b) and (c) are typical episodes observed in the process of system evolution. In the episode plotted in Fig.~\ref{fig2}(b), the $N=10$ trajectories are well separated from each other, signifying the desynchronization nature of the neurons. Different from this, in the episode plotted in Fig.~\ref{fig2}(c), it is seen that the trajectories of neurons $4$, $6$ and $9$ are completely overlapped, and there are only $8$ distinct trajectories in total. That is, the subset of neurons, $V_c=\{4,6,9\}$, are synchronized and form a cluster temporally during this episode. A close look at Fig.~\ref{fig2}(c) also shows that except for the three neurons, the other neurons are kept desynchronized. To have a global picture on the synchronization behavior of the cluster, we plot in Fig.~\ref{fig2}(d) the time evolution of cluster-based synchronization error, $\delta x_{clu}(t)=\sum_l|x_l(t)-\hat{x}(t)|/3$, with $l\in V_c$ and $\hat{x}(t)=\sum_l x_l(t)/3$ the averaged membrane potential. We see that $\delta x_{clu}$ is staying around $0$ (below $0.1$) for most of the time, but is jumping to large values ($\sim 1.0$) sporadically, showing the typical feature of on-off intermittency observed in chaotic systems~\cite{Intermit:platt1993,Intermit:wxg2009}. 

It is worth mentioning that the intermittent cluster synchronization depicted by $\delta x_{clu}$ in Fig.~\ref{fig2}(d) is essentially different from the desynchronization behaviors of the global network revealed by $\delta x_{net}$ in Fig.~\ref{fig2}(a). One difference lies in the frequency of the desynchronization bursts. Compared to the results of global synchronization, the frequency of cluster desynchronization is slower by several orders of magnitude. Another difference lies in the minimum synchronization error. In Fig.~\ref{fig2}(a), the minimum value of $\delta x_{net}$ is about $0.1$; in Fig.~\ref{fig2}(d), the minimum value of $\delta x_{clu}$ is close to $0$ (smaller than $1\times 10^{-5}$). Numerical results thus portray the following picture of network evolution: embedded in the background of desynchronized neurons, a small-size neuron cluster is switching between the synchrony and asynchrony states in an intermittent fashion. This phenomenon, which is named cluster breathing in the present work, is reminiscent of the slow-rhythm oscillations~\cite{Singer1999,BrainWebSyn,Ward2003} and spontaneous state switching observed in the neocortex~\cite{Poskanzer:Switching2016,UpdtateAstro2011}. 

\begin{figure}[tbp]
\begin{center}
\includegraphics[width=0.7\linewidth]{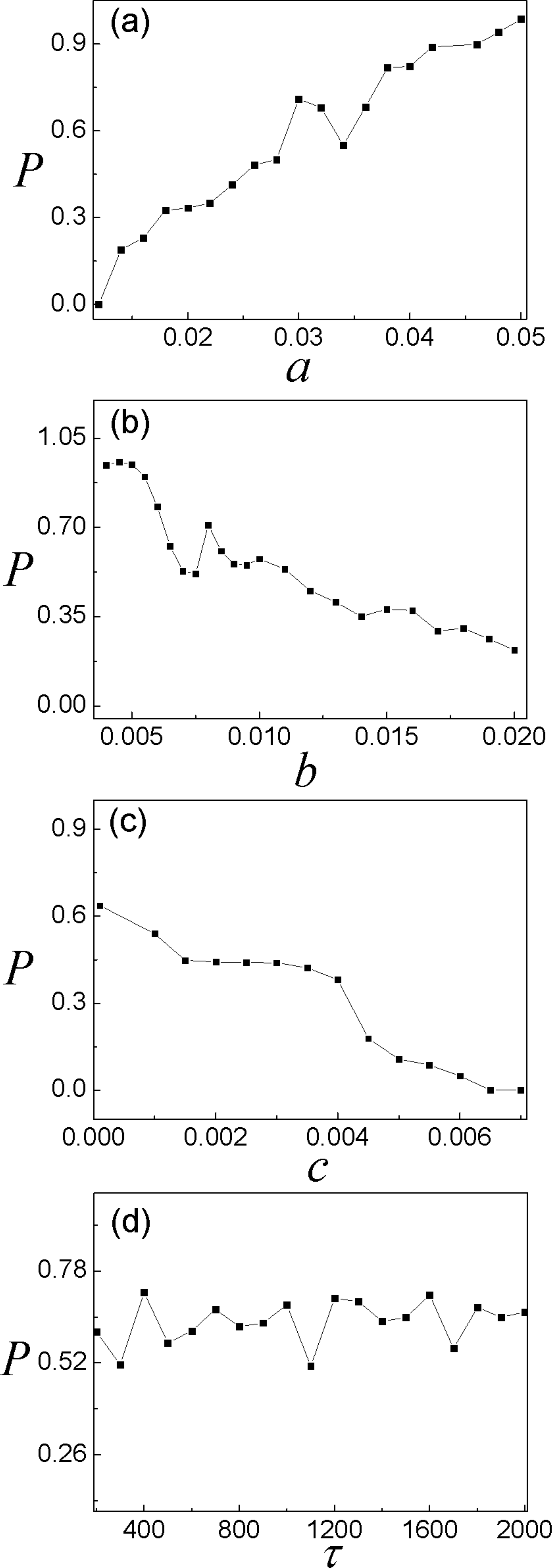}
\caption{Influences of astrocyte parameters on cluster breathing. $P$ is the fraction of time the cluster-based synchronization error $\delta x_{clu}$ exceeds the threshold $0.1$. (a) $P$ versus $a$ (the attenuation coefficient). (b) $P$ versus $b$ (the generic coupling strength). (c) $P$ versus $c$ (the constant input). (d) $P$ versus $\tau$ (the time-delay coefficient). Each result is averaged over a time period of $T=1\times 10^5$.} 
\label{fig3}
\end{center}
\end{figure}

As in our model the regulating effect of astrocyte on neural activities is realized by modifying the synaptic strength, $\varepsilon(t)$, it is natural to conjecture that the slow-frequency breathing of the cluster may be accompanied by a slow variation of the coupling strength. To check this out, we plot in Fig.~\ref{fig2}(e) the time evolution of $\varepsilon$ (which is also the variable denoting the glutamate concentration in the astrocyte). It is seen that $\varepsilon$ is oscillating randomly within the range $(0.1,0.3)$ at a frequency much higher than that of cluster breathing. [As a matter of fact, as the evolution of $\varepsilon$ is driven by the time-delayed synchronization order parameter, $R(t-\tau)$, in a linear fashion, the oscillating frequency of $\varepsilon$ is similar to the frequency of the network desynchronization depicted in Fig.~\ref{fig2}(a).] This unexpected result triggers our interest about the factors influencing the frequency of cluster breathing. To investigate, we change the parameters of the astrocyte, $(a,b,c,\tau)$, and check their impacts on the breathing frequency individually. Here, breathing frequency is defined as the fraction of time, $p$, that the cluster-based synchronization error, $\delta x_{clu}$, exceeds the threshold error $\delta x^c_{clu}=0.1$. The results are plotted in Fig.~\ref{fig3}. We see that $p$ is increased with $a$ (the attenuation coefficient), but is decreased with $b$ (the generic coupling strength) and $c$ (the constant input). The numerical results also suggest that $p$ is not influenced by $\tau$ (the time delay coefficient), as depicted in Fig.\ref{fig3}(d). 

As the strength of the neural synapses is updated according to the network synchronization order parameter [see Eq.~(\ref{glu})], the neuron-astrocyte network model can be treated as an adaptive complex network (specifically, the strategy of global network adaption)~\cite{AdpNetRev:Ghosh2022}. This treatment makes it plausible to interpret the breathing phenomenon from the perspective of adaptive synchronization. However, in previous studies of adaptive network synchronization, a common finding is that if the coupling strength is updated by the synchronization order parameter in a linear fashion [such as the one described by Eq.~(\ref{glu})], the network will be finally settled onto a stationary state of fixed coupling strength and constant synchronization order parameter~\cite{AdpNet:PDL2008}, instead of a non-stationary synchronization state of fluctuating synchronization order parameter. For our model of complex neuron-astrocyte network, if the network is settled onto a stationary state, the synchronization order parameter will be approximately a constant $R_0$. According to Eq.~(\ref{glu}), in such a case the coupling strength (the astrocyte glutamate concentration) will be stabilized onto the equilibrium point $\varepsilon_0=(bR_0+c)/a$. Yet the numerical results in Fig.~\ref{fig2} show that, in contrast to the expectations, both the synchronization order parameter and the coupling strength are oscillating wildly with time. Furthermore, the results in Fig.~\ref{fig3} show that the non-stationary feature of the neural network is dependent on the astrocyte parameters, signifying again the oscillating nature of the network dynamics. Hence, the conventional methods developed for adaptive networks are not suitable for interpreting the non-stationary dynamics of the neuron-astrocyte network and, to explore the breathing phenomenon, new methods should be adopted. As cluster synchronization is overlooked in previous studies of adaptive network synchronization, one might speculate that the non-stationary feature might be attributed to cluster synchronization. If this is the case, then why the cluster is also non-stationary, i.e., presenting the breathing phenomenon, and what is the mechanism generating the breathing? Interested in these questions, we next conduct a theoretical analysis on the dynamical properties of the synchronization cluster.    

\section{Theoretical analysis}

As the neurons are coupled through chemical synapses and the structure of the neural network is irregular (the degrees of the neurons are not identical), it is difficult to achieve global network synchronization. However, if there is a subset of neurons whose incoming signals are identical, it is possible for this subset of neurons to be synchronized. One extreme case is the synchronization of isolated chaotic oscillators subjected to the common noise, i.e., the phenomenon of noise-induced chaos synchronization~\cite{NoiseSyn:KM,NoiseSyn:CZ}. Therefore, for a subset of neurons to be synchronizable, the necessary condition is that they are enclosed by the same set of neighbors on the network, or, speaking alternatively, the neurons in the subset are symmetric with each other in the sense that their permutations do not change the network dynamics. A check of the network structure plotted in Fig.~\ref{fig1}(a) does show this feature. To be specific, the subset of neurons $V_c=\{4,6,9\}$ (which form the cluster) are all coupled to neurons $2$, $5$ and $7$. (For small-size complex networks, the symmetric neurons can be identified visually. For large-size complex networks, the symmetric neurons can be identified by sophisticated techniques such as the one developed from computational group theory~\cite{Pecora2014} or the method of structural position vector~\cite{NetworkSym:LYS2022}.)  

\begin{figure}[tbp]
\begin{center}
\includegraphics[width=0.75\linewidth]{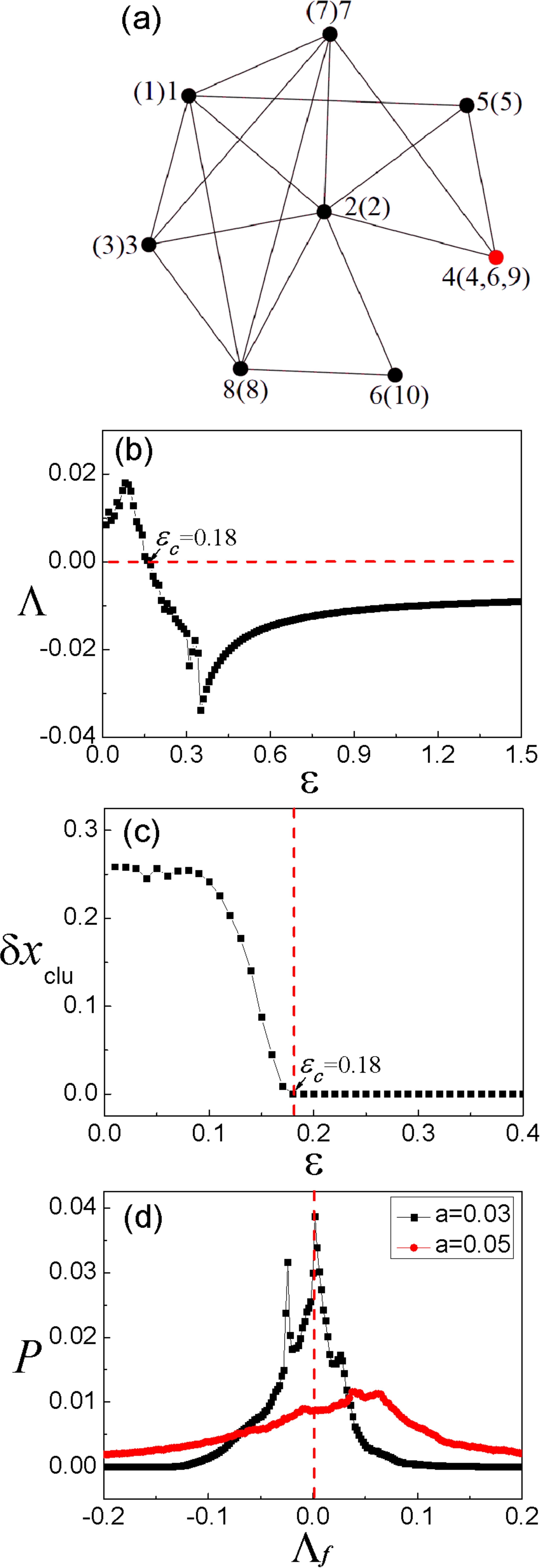}
\caption{(Color online) (a) The simplified network governing the dynamics of the cluster synchronization state. Node $4$ (red) denotes the symmetric cluster (neurons 4, 6 and 9) in the original network. Nodes 1,2,3,5,6,7 and 8 correspond to neurons 1,2,3,5,10,7 and 8 in the original network, respectively. (b) For the scenario of constant coupling strength, the variation of the largest conditional Lyapunov exponent, $\Lambda$, with respect to $\varepsilon$. $\Lambda<0$ for $\varepsilon>\varepsilon_c\approx 0.18$. (c) For constant couplings, the variation of cluster-based synchronization error, $\delta x_{clu}$, with respect to $\varepsilon$. (d) The probability distribution of the finite-time Lyapunov (FLE) exponent, $\Lambda_{f}$, for the astrocyte parameters $a=3\times 10^{-2}$ (black) and $a=5\times 10^{-2}$ (red).} 
\label{fig4}
\end{center}
\end{figure}

Having identified the set of symmetric neurons forming potentially a synchronization cluster, we next analyze the stability of the cluster. When the symmetric neurons are completely synchronized, the system dynamics is captured by a simplified network whose dynamics reads (obtained by replacing the symmetric neurons with a virtual neuron)
\begin{equation}
\dot{\bm{s}}_{i'} = \bm{F}(\bm{s}_{i'}) + \varepsilon(t)\sum_{j'=1}^{N'}c_{i'j'}\bm{H}(\bm{s}_{i'},\bm{s}_{j'}),\\
\label{quonet}
\end{equation}
where $i', j'=1,\dots,N'$ are the new neuron indices, $N'=N-m+1$ is the size of the simplified network, and $m$ is the number of symmetric neurons in the cluster. For the specific neural network plotted in Fig.~\ref{fig1}(a), we have $m=3$ and $N'=8$. The structure of the simplified network is plotted in Fig.~\ref{fig4}(a), in which node $4$ (colored in red) is the virtual neuron representing the symmetric neurons in the original network. Except for the virtual node, there is a one-to-one correspondence between nodes in the simplified network and neurons in the original network, $i'\rightarrow i$, as depicted in Fig.~\ref{fig4}(a). In Eq.~(\ref{quonet}), $\bm{s}_{i'}$ stands for the state vector of neuron $i'$, $\bm{F}$ denotes the dynamics of isolated neuron (the HR oscillator), and $\bm{H}(\bm{s}_{i'},\bm{s}_{j'})=[(V_{r}-s_{i'})h(s_{j'}),0,0]^T$ represents the coupling function (chemical synapses). The coupling matrix of the simplified network, $\bm{C}$, is constructed from the adjacency matrix $\bm{A}$ according to the network symmetry: $c_{i'j'}=a_{ij}$ for $j'\neq 4$, and $c_{i'4}=\sum_{l}a_{il}$ with $l\in V_c$ (the set of symmetric neurons). (Please see Supplementary Material for more details.) Equation~(\ref{quonet}) governs the dynamics of the cluster synchronization state and, importantly, defines the synchronization manifold of the symmetric neurons, i.e., $\bm{s}_{4}$, though the cluster might be unstable. 
 
To investigate the synchronizability of the cluster, we introduce small random perturbations to the cluster synchronization state defined by Eq.~(\ref{quonet}) and check the time evolution of the perturbations. Let $\delta \bm{x}_i=\bm{x}_i-\bm{s}_{i'}$ be an infinitesimal random perturbation added onto neuron $i$ (note that the reference states of the symmetric neurons are identical), the evolution of the perturbations is governed by the following set of variational equations [obtained by linearizing Eq.~(\ref{HR}) around the synchronization state defined by Eq.~(\ref{quonet})]
\begin{equation}
\begin{aligned}
\delta \dot{\bm{x}}_i=&D\bm{F}(\bm{s}_{i'})\delta \bm{x}_i-\varepsilon(t)\sum^N_{j=1}a_{ij}\bm{H}(\bm{s}_{j'})\delta \bm{x}_{i}\\
&+\varepsilon(t)(V_r-s_{i'})\sum^N_{j=1}a_{ij}D\bm{H}(\bm{s}_{j'})\delta \bm{x}_{j},
\end{aligned}
\label{vareq}
\end{equation}     
with $D\bm{F}$ and $D\bm{H}$ being the Jacobian matrices. For the symmetric neurons, the variational equations can be rewritten as
\begin{equation}
\begin{aligned}
\delta \dot{\bm{x}}_l&=[D\bm{F}(\mathscr{S})-\varepsilon(t)\sum^N_{j=1}a_{lj}\bm{H}(\bm{s}_{j'})]\delta \bm{x}_l\\
&+\varepsilon(t)(V_r-\mathscr{S})\sum_{l'\in V_c}a_{ll'}D\bm{H}(\mathscr{S})\delta \bm{x}_{l'}+\bm{I}(t),\\
\end{aligned}
\label{vareq2}
\end{equation}  
with $l, l'\in V_c$, $\mathscr{S}=\bm{s}_4$ is the synchronization manifold of the cluster, and $\bm{I}(t)=\varepsilon(V_r-\mathscr{S})\sum_{j\notin V_c}a_{ij}D\bm{H}(\bm{s}_{j'})\delta \bm{x}_{j}$ is the common signal received by the symmetric neurons. Defining $\delta \bm{y}_l\equiv\delta \bm{x}_l-\left<\delta \bm{x}_c\right>$, with $\left<\delta \bm{x}_c\right>=\sum_{l\in V_c} \delta \bm{x}_l/m$ the averaged perturbations of the symmetric neurons, we have
\begin{equation}
\begin{aligned}
\delta \dot{\bm{y}}_l=&[D\bm{F}(\mathscr{S})-\varepsilon(t)\sum^N_{j=1}a_{lj}\bm{H}(\bm{s}_{j'})]\delta \bm{y}_l\\
&+\varepsilon(t)(V_r-\mathscr{S})D\bm{H}(\mathscr{S})\sum_{l'\in V_c}a_{ll'}\delta \bm{y}_{l'}.\\
\end{aligned}
\label{vareq3}
\end{equation}   
Transforming Eq.~(\ref{vareq3}) into the mode space spanned by the eigenvectors of the submatrix $\tilde{A}=\{a_{ll'}\}_{m\times m}$, we have 
\begin{equation}
\begin{aligned}
\delta \dot{\bm{z}}_k=&[D\bm{F}(\mathscr{S})-\varepsilon(t)\sum^N_{j=1}a_{lj}\bm{H}(\bm{s}_{j'})]\delta \bm{z}_k\\
&+\varepsilon(t)\lambda_k(V_r-\mathscr{S})D\bm{H}(\mathscr{S})\delta \bm{z}_{k},\\
\end{aligned}
\label{msf}
\end{equation}   
with $k=1,\ldots,m$ the mode index, $\delta\bm{z}_k$ the $k$th mode, and $\lambda_1\geq\ldots\geq\lambda_m$ the ordered eigenvalues of $\tilde{A}$. Let $\Lambda_k$ be the largest Lyapunov exponent of the $k$th mode calculated from Eq.~(\ref{msf}), the necessary condition for the cluster to be synchronizable is $\Lambda_k<0$ for modes $k=2,\ldots,m$. (The mode associated with $\lambda_1$ is in parallel to the synchronization manifold $\mathscr{S}$.) Equation~(\ref{msf}) is our main theoretical result, which tells that the synchronizability of the cluster is jointly determined by the cluster synchronization state, $\{\bm{s}_{i'}\}$ (with $i'=1,\ldots,N'$) and the couplings between the symmetric neurons, $\tilde{A}$.

Before utilizing Eq.~(\ref{msf}) to explore the mechanism of cluster breathing, we first validate the theoretical prediction by considering the scenario of constant couplings. For the network shown in Fig.~\ref{fig1}(a), as there is no connection between the symmetric neurons, all elements in $\tilde{A}$ are zero and the eigenvalues are therefore null. Setting $\lambda=0$ in Eq.~(\ref{msf}), we have
\begin{equation}
\delta \dot{\bm{z}}=[D\bm{F}(\mathscr{S})-\varepsilon\sum^{N'-1}_{j'=1}c_{4j'}\bm{H}(\bm{s}_{j'})]\delta \bm{z}, \\
\label{msf1}
\end{equation}  
with $j'\in \{2,5,7\}$ neurons connected to the virtual neuron [node colored in red in Fig.~\ref{fig4}(a)] in the simplified network. Solving Eqs.~(\ref{quonet}) and (\ref{msf1}) together numerically, we are able to find the variation of the largest Lyapunov exponent of the perturbation modes, $\Lambda$, with respect to $\varepsilon$. The results are plotted in Fig.~\ref{fig4}(b). We see that $\Lambda<0$ when $\varepsilon>\varepsilon_c\approx 0.18$. That is, for the situation of constant couplings, the theory predicts that the cluster will be synchronized when $\varepsilon>\varepsilon_c$. To validate the theoretical prediction, we set $\varepsilon$ as constant in Eqs.~(\ref{HR}) and (\ref{coupling}), and conduct a direct simulation for the system evolution. Shown in Fig.~\ref{fig4}(c) is the variation of the cluster-based synchronization error, $\delta x_{clu}$, with respect to $\varepsilon$. We see that $\delta x_{clu}$ crosses $0$ at about $\varepsilon_c\approx 0.18$, which is in good agreement with the theoretical prediction. 

We proceed to analyze the mechanism of cluster breathing for the scenario of adaptive coupling described by Eq.~(\ref{glu}). As on-off intermittency is typically observed in the vicinity of the bifurcation point of the system dynamics~\cite{Intermit:platt1993,Intermit:wxg2009}, a natural expectation will be that with the breathing of the cluster, the coupling strength is oscillating around $\varepsilon_c$ with a similar frequency. However, a check of the numerical results plotted in Fig.~\ref{fig2}(e) shows that, while $\varepsilon$ does oscillate around the critical coupling $\varepsilon_c$ (shown as the red-dashed line), the oscillating frequency is much higher than the frequency of cluster breathing. Noticing that in Eq.~(\ref{glu}) the evolution of $\varepsilon(t)$ is driven by the synchronization order parameter, $R(t-\tau)$, in a linear fashion, the oscillation of $\varepsilon(t)$ thus is naturally correlated to the oscillation of $R(t)$. (Numerical result shows that the Pearson correlation coefficient between the two signals is about $0.7$). Numerical results thus suggest that cluster breathing is not attributed to the oscillation of the coupling strength. [It is worth mentioning that while random oscillation of $R(t)$ is typically observed at the bifurcation point of global network synchronization~\cite{Intermit:wxg2009}, the oscillation amplitude is usually very small. Here, with the breathing of the cluster, the value of $R(t)$ is oscillating widely between $0$ (the complete desynchronization state) and $1$ (the global synchronization state)].     

We next analyze the breathing activities of the cluster from the point of view of finite-time Lyapunov exponent (FLE)~\cite{FLE:Pikovsky1995,FLE:LSS2018}. FLE is defined as 
\begin{equation}
\Lambda_f=\frac{1}{\Delta T}\ln \left| {\bm{Q}}(\Delta T)\cdot \bm{u}_0 \right|, \\ 
\label{fle}
\end{equation}
with $\Delta T$ being the time interval over which FLE is averaged, $\bm{Q}(\Delta T)$ being the matrix solution of Eq.~(\ref{msf1}), and $\bm{u}_0$ being the random unit vector in the tangent space of the synchronous manifold associated with the symmetric cluster. The cluster is temporally (locally) stable if $\Lambda_f<0$, and it is temporally unstable if $\Lambda_f>0$. For a stable cluster, while the largest conditional Lyapunov exponent is negative, the value of $\Lambda_f$ could be positive for some time intervals. These intervals correspond to regions in the tangent space in which the infinitesimal perturbations diverge from each other, giving rise to bursts of cluster desynchronization. The asymptotic exponent, i.e., the largest conditional Lyapunov exponent, $\Lambda$, is just the weighted sum of the temporally positive exponents when the trajectory visits the expanding regions ($\Lambda_f>0$) and the temporally negative exponents when the trajectory visits the contracting regions ($\Lambda_f<0$). For the breathing state shown in Fig.~\ref{fig2}, the largest conditional Lyapunov exponent calculated from Eq.~(\ref{msf1}) is about $\Lambda=-1\times 10^{-3}$. That is, the negative FLE exponents prevail over the positive exponents and, as the consequence, the cluster is asymptotically stable. Setting $\Delta T=10$ (which is approximately the time interval between successive spikes), we plot in Fig.~\ref{fig4}(d) the probability distribution of $\Lambda_f$. The results show that, while $\Lambda_f$ is mostly distributed in the negative (stable) regime, the distribution has a distinct peak (around $\Lambda_f= 2\times 10^{-2}$) in the positive (unstable) regime. It is just this peak that generates the desynchronization bursts of the cluster. Specifically, most of the time the system dynamics is evolving in the contracting regions of negative $\Lambda_f$ in the phase space, during which the trajectories of the symmetric neurons are staying close to each other and $\delta x_{clu}$ is around $0$; occasionally, the system dynamics enter the expanding regions of positive $\Lambda_f$, during which noise perturbations induce a quick divergence of the trajectories, resulting in a sudden increase of $\delta x_{clu}$. Therefore, from the point of view of FLE, the breathing activities can be interpreted as the random itinerancy of the system dynamics between the contracting and expanding regions in the tangent space of the cluster synchronization state.     

Having revealed the mechanism of cluster breathing, it is straightforward to anticipate that cluster breathing can be also observed when the largest conditional Lyapunov exponent of the perturbation mode is positive, i.e., $\Lambda>0$. In this case, though the cluster is globally unstable, there are time intervals with negative $\Lambda_f$, during which the trajectories of the symmetric neurons are contracted to the vicinity of the synchronization manifold, making $\delta x_{clu}$ stay around $0$. However, once the trajectories visit the expanding regions with positive $\Lambda_f$, they will be quickly diverged from each other, leading to the generation of a desynchronization burst. To show an example, we set $a=5\times 10^{-2}$ while keeping the other parameters identical to Fig.~\ref{fig2}. We simulate again the network dynamics and calculate the largest conditional Lyapunov exponent of the cluster according to Eq.~(\ref{msf1}). The numerical result shows that $\Lambda=1.2\times 10^{-2}$. As such, the cluster is globally unstable in this case. Plotted in Fig.~\ref{fig4}(d) is the distribution of the FLE for $\Delta T=10$. We see that while $\Lambda_f$ is mostly distributed in the positive (unstable) regime, a significant fraction of $\Lambda_f$ is located in the negative (stable) regime, suggesting that the cluster is also breathing with time. This prediction is verified by numerical simulations (please see Supplementary Material for the breathing activities of the cluster). 

Comparing the FLE distributions of $a=3\times 10^{-2}$ ($\Lambda<0$) and $a=5\times 10^{-2}$ ($\Lambda>0$) in Fig.~\ref{fig4}(d), it is noticed that the fraction of positive $\Lambda_f$ in the latter is significantly larger to that of the former. That is, compared to the breathing state generated by the parameter $a=3\times 10^{-2}$, the breathing state generated by the parameter $a=5\times 10^{-2}$ has a higher probability of visiting the expanding regions in the phase space tangent to the cluster synchronization manifold. As desynchronization bursts are generated in the expanding regions, the latter therefore has a higher breathing frequency as compared with the former. This explains the numerical results presented in Fig.~\ref{fig3}(a), in which the breathing frequency, $p$, is shown to be increased with the astrocyte parameter $a$ monotonically. The increase of $p$ with $a$ can be also interpreted by the dependence of the equilibrium point $\varepsilon_0=(bR_0+c)/a$ on the astrocyte parameters $(a,b,c)$. With the increase of $a$, the value of $\varepsilon_0$ will be decreased. As $\varepsilon(t)$ is oscillating around $\varepsilon_0$, according to the results plotted in Fig.~\ref{fig4}(b), the trajectories have a higher probability of entering the unstable regime ($\Lambda>0$), resulting in the increased breathing frequency. The influences of the parameters $b$, $c$, and $\tau$ on the breathing frequency can be understood in a similar way. Specifically, by increasing $b$ or $c$, the equilibrium point $\varepsilon_0$ will be increased and the probability of visiting the stable regime ($\Lambda<0$) will be increased, leading to the decreased breathing frequency [see Figs.~\ref{fig3}(b) and (c)]. As $\varepsilon_0$ is independent of $\tau$, the breathing frequency is not influenced by the time-delay coefficient. This explains the numerical results plotted in Fig.~\ref{fig3}(d).  

\section{Generalization}

\begin{figure}[tbp]
\begin{center}
\includegraphics[width=0.95\columnwidth]{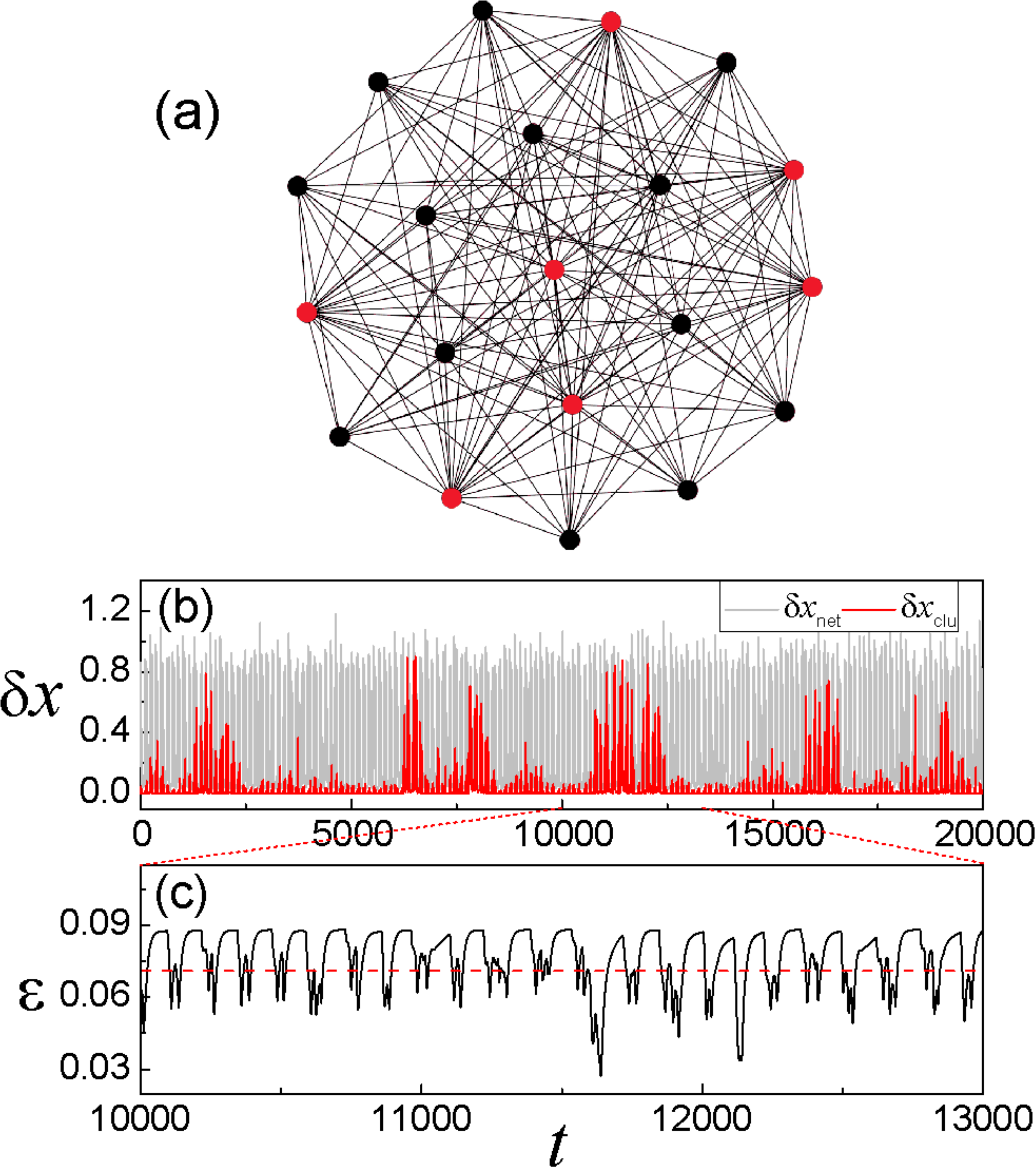}
\caption{(Color online) (a) Cluster breathing in a larger-size complex network containing $N=20$ neurons. (a) The network structure. The symmetric neurons are colored in red. (b) Time evolutions of the network synchronization error (grey color), $\delta x_{net}$, and the cluster-based synchronization error (red color), $\delta x_{clu}$. (c) Time evolution of the coupling strength, $\varepsilon(t)$. Red dashed line denotes the critical constant coupling, $\varepsilon_c\approx 0.07$, for synchronizing the cluster.}
\label{fig5}
\end{center}
\end{figure}

We finally check the generality of the phenomenon of cluster breathing by considering other neuron-astrocyte network models, including increasing the network size, adopting different neural dynamics, and changing the synaptic coupling function. Shown in Fig.~\ref{fig5}(a) is a random network consisting of $N=20$ neurons, which is generated by removing randomly $50$ links from a globally connected network of the same size. Adopting this network structure and setting $(a,b,c)=(0.11,8\times 10^{-3},5\times 10^{-4})$ as the astrocyte parameters, we check again the system evolution by simulations. The time evolution of the network synchronization error, $\delta x_{net}$, is plotted in Fig.~\ref{fig5}(b) (the grey color). The wild oscillation of $\delta x_{net}$ implies that the neurons are desynchronized. By the method of structural position vector~\cite{NetworkSym:LYS2022}, we are able to find the set of symmetric neurons on the network, $V_c=\{14,15,16,17,18,19,20\}$, which are colored in red in Fig. ~\ref{fig5}(a). The time evolution of the cluster-based synchronization error, $\delta x_{clu}$, is also plotted in Fig.~\ref{fig5}(b) (the red color). It is seen that the cluster undergoes the typical process of on-off intermittency. That is, the cluster is breathing with time. Shown in Fig.~\ref{fig5}(c) is the time evolution of the coupling strength, $\varepsilon(t)$. We see that, similar to the results presented in Fig.~\ref{fig2}(e), the coupling strength is oscillating at a frequency much higher than the frequency of cluster breathing. Following the procedures presented in Sec. IV, we conduct a theoretical analysis on the stability of the cluster, which shows that the largest conditional Lyapunov exponent of the cluster is about $\Lambda=-2.5\times 10^{-2}$. Further analysis also shows that in the scenario of constant coupling, the cluster is synchronizable when $\varepsilon>\varepsilon_c\approx 0.07$, which is denoted by the red-dashed line in Fig.~\ref{fig5}(c). (Please see Supplementary Material for coupling matrices of the original and simplified networks.) 

\begin{figure}[tbp]
\begin{center}
\includegraphics[width=0.9\columnwidth]{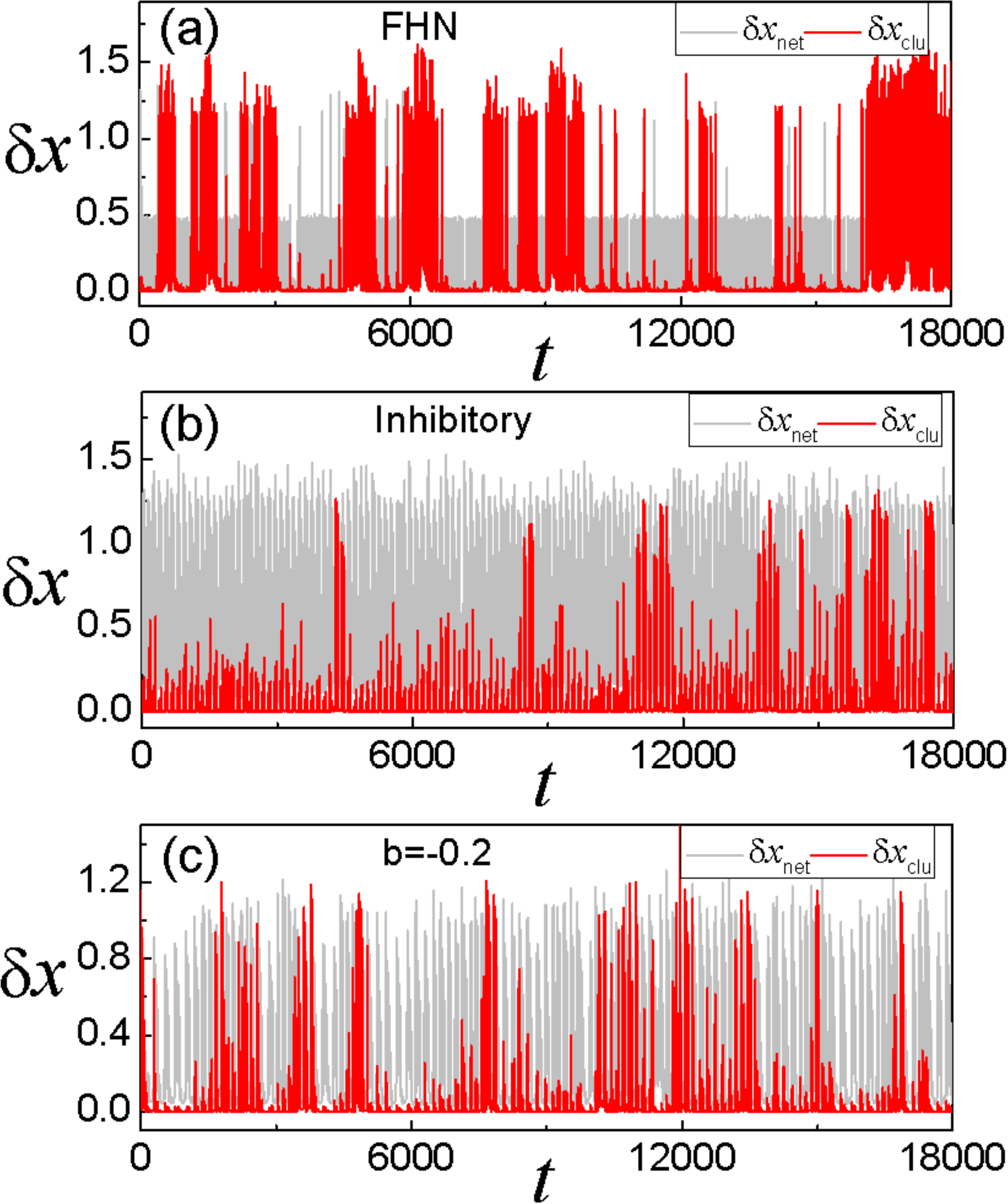}
\caption{(Color online) For the network plotted in Fig.~\ref{fig1}(a), the synchronization behaviors for (a) coupled FHN oscillators with excitatory chemical synapses, (b) coupled HR oscillators with inhibitory chemical synapses, and (c) inhibitory neuron-astrocyte coupling $b=-0.2$. Shown in each panel are the time evolutions of the network synchronization error (grey color), $\delta x_{net}$, and the cluster-based synchronization error (red color), $\delta x_{clu}$.}
\label{fig6}
\end{center}
\end{figure}

Cluster breathing is also observed when other types of neural oscillators are adopted as the local dynamics. To show an example, we adopt the FitzHugh-Nagumo (FHN) oscillator as the neural dynamics and, by the same network studied in Fig.~\ref{fig2} ($N=10$ and $V_c=\{4,6,9\}$), check again the synchronization behaviors of the neurons. The dynamics of the FHN oscillator in the isolated form is governed by the equations~\cite{FHN:Scholl2013}: $(dx/dt, dy/dt)^T$ = $[c'(x-x^{3}/3+y+I'_{ext}), -1/c'(x-a'+b'y)]^T$. The parameters of the FHN oscillators are chosen as $(a',b',c',I'_{ext})=(0.7,0.8,3.0,-0.4)$, by which the oscillators are oscillatory. Still, the oscillators are coupled by the excitatory chemical synapses. Setting the astrocyte parameters as $(a,b,c)=(1,9\times 10^{-2},5\times 10^{-3})$, we plot in Fig.~\ref{fig6}(a) the time evolution of the network synchronization error (the grey color), $\delta x_{net}$, and the cluster-based synchronization error (the red color), $\delta x_{clu}$. Apparently, the network is desynchronized while the cluster is breathing with time. To check the impact of the coupling function on cluster breathing, we keep the other settings of the system unchanged but set $V_r=-2$ in Eq.~(\ref{coupling}) for the chemical synapses. As $V_r<0$, the synapses are inhibitory. By the astrocyte parameters $(a,b,c)=(0.01,0.01,1\times 10^{-3})$, we plot in Fig.~\ref{fig6}(b) the time evolutions of the network synchronization error, $\delta x_{net}$, and the cluster-based synchronization error, $\delta x_{clu}$. Again, we see that embedded in the desynchronization background, a cluster is switching slowly between the synchrony and asynchrony states in an intermittent fashion.

Besides the models tested above, cluster breathing is also observed in other types of neuron-astrocyte network models, including complex networks with inhibitory neuron-astrocyte couplings [i.e., $b<0$ in Eq.~(\ref{glu})], complex networks of coupled Hodgkin-Huxley (HH) oscillators, and noise-free systems. Shown in Fig.~\ref{fig6}(c) are the results for inhibitory neuron-astrocyte coupling $b=-0.2$ (the other settings are identical to the network model studied in Fig.~\ref{fig2}), which are similar to the breathing activities generated by the excitatory coupling as plotted in Fig.~\ref{fig2}(d). (Please see Supplementary Material for the results of coupled HH oscillators and noise-free systems.) These additional results suggest that cluster breathing is a general phenomenon for complex neuron-astrocyte networks. 

\section{Discussions and conclusion}

The current study sheds light on the molecular and cellular mechanisms of astrocytes in regulating the firing activities of neurons, and gives insights into the spontaneous state switching in the neocortex. Whereas neuroscience is primarily a neuronal field, accumulating evidence shows that glial cells participate actively in a variety of brain functions~\cite{Glia:Barres,Glia:VA}. These studies, however, are mostly based on experiments, while few mathematical models have been proposed for interpreting the experimental observations. In particular, recent {\it in vivo} experiments show that astrocytes, the most abundant glial cells in the brain, play an important role in modulating the synchronization activities of cortical neurons~\cite{UpdtateAstro2011,Poskanzer:Switching2016}, yet the signaling mechanism remains not clear. Specifically, the experimental results suggest that the regulating effects of astrocytes on neuronal synchronization activities (e.g., triggering the emergence of UP state by stimulating a single astrocyte) are realized by the processes of glutamate uptake and release, but it remains elusive how the transient increase of extracellular glutamate promotes the UP state and how the cortical state is spontaneously switching between different synchronization states at a slow frequency. In our present work, treating astrocyte as a glutamate reservoir accessing to the neighboring neurons in a local brain area, we propose the model of complex neuron-astrocyte network and, by numerical simulations and theoretical analysis, investigate the interplay between astrocyte and neural network. The new network model reproduces successfully not only the emergence of cortical UP states (the cluster synchronization state), but also the slow-switching of the cortical state (between cluster synchronization and cluster desynchronization states) as reported in experimental studies~\cite{UpdtateAstro2011,Poskanzer:Switching2016}. More importantly, our theoretical analysis of the stability of the cluster synchronization state reveals that the emergence the UP state is attributed to the temporally enhanced neural couplings and the slow-switching of the cortical state is due to the local stability of the cluster synchronization manifold. These findings support the general idea that glutamate regulation is the signaling mechanism for cortical state switching~\cite{Rev:astrocyte2021}, and provide a microscopic picture for the mutual interactions between the astrocytic and neuronal dynamics. 

The findings in the current study enrich our understanding on the cluster synchronization behaviors of complex networks. Though cluster synchronization has been observed in a variety of complex networks and some criteria for generating synchronization clusters have been obtained, the clusters reported in the literature are mostly stationary~\cite{SynSymtry:Russo2011,CS:VN2013,Pecora2014,LWJ-1,FS:2016,LWJ-2,YC:2017,CS:WYF2019}. That is, once the cluster is synchronized, oscillators inside the cluster will be kept as synchronized during the process of network evolution. The stationary feature of cluster synchronization is observed not only in static complex networks of fixed coupling strength, but also in evolutionary networks in which the strength between coupled oscillators is updated adaptively according to the synchronization degree~\cite{AdpNet:Scholl2021}. Our present work shows that by incorporating astrocytes into the conventional model of neural networks, the interplay between the astrocyte and the neural network leads to naturally a slow switching of the cluster between the synchrony and asynchrony states. Furthermore, the results of FLE analysis suggest that the slow state switching is a joint effect of the oscillation of the neural coupling strength (modified by the synchronization order parameter) and the random itinerancy of the neural dynamics (between the contracting and expanding regions) in the phase space. For the significance of slow-wave oscillations to a wide range of cognitive functions (e.g., sleep and memory), the phenomenon of cluster breathing and the breathing mechanism we have revealed might have broad implications. [A phenomenon similar to cluster breathing is chimera breathing, which is typically observed in regular networks of coupled phase oscillators~\cite{ChimeraBreathing:Abrams2008,ChimeraBreathing:Laing2009,ChimeraBreathing:MIB:2017}. In chimera breathing, the size of the coherent domain consisting of synchronized oscillators is timely changing, leading to the oscillation of the synchronization order parameter. Different from chimera breathing in which the synchronization relationship of the oscillators is defined on phase (namely phase synchronization or frequency locking) and the synchronization order parameters of the whole network is oscillating periodically with time, in cluster breathing the cluster is defined on complete synchronization and the cluster is switching between the synchrony and asynchrony states in an intermittent fashion. In addition, the mechanism of chimera breathing is also different from that of cluster breathing~\cite{ChimeraBreathing:Abrams2008,ChimeraBreathing:Laing2009,ChimeraBreathing:MIB:2017}.]

The neuron-astrocyte network model proposed in the current study can be improved in several aspects. Firstly, for simplicity and demonstration purpose (to highlight the fact that cluster breathing is an intrinsic property of the system dynamics), we have treated the astrocyte as a glutamate reservoir that responds linearly to the synchronization order parameter. In realistic situations, the processes of glutamate uptake and release involve a number of variables, e.g., Ca$^{2+}$ and IP$_{3}$, and the dynamics of the variables are governed by sophisticated nonlinear equations~\cite{AstroModel:Young1992,AstroModel:LYX1994,AstroModel:UG:2006}. To better describe the regulating effects of astrocyte on neurons and to make the studies more relevant to the experimental results, sophisticated nonlinear models should be adopted to describe the dynamics of astrocyte. Secondly, the model we have proposed consists of a single astrocyte and the astrocyte is assumed to be coupled with all neurons on the network, which are different from the realistic situations in which a large number of astrocytes and neurons are coupled in a complex fashion. Experimental evidences suggest that, like neurons, astrocytes are also densely connected (through gap junctions) as a complex network, and each astrocyte could contact up to thousands of neurons. A more accurate model of the neocortex therefore should include at least two interacting complex networks, one for astrocytes and the other one for neurons. Thirdly, in our studies the neural network is constructed by links (chemical synapses) of uniform coupling strength and the neurons are of identical dynamics, which are also different from the realistic situations in which synapses are weighted and directed and the neurons are of different dynamics. The generalization of the findings to the general neural networks will be important and necessary. (In the general network, as rigorous network symmetry does not exist, cluster synchronization should be redefined, e.g, based on phase synchronization~\cite{BrainWebSyn}.) Finally, the neural networks studied in our present work contain only one symmetric cluster, while the cognitive functions of the brain are normally supported by dynamical patterns of spatially distributed synchronization clusters~\cite{NetworkSynBrain,NetNeurosci2017}. By considering large-scale neural networks, it might be possible to observe the breathing of multiple clusters. If this phenomenon indeed is observable, then an interesting question will be how the breathing of the different clusters is coordinated by astrocytes. We hope these issues could be addressed by further studies.    

To summarize, inspired by recent experimental studies on the regulation of neuronal synchronization activities by astrocytes, we have proposed a complex neuron-astrocyte network model and investigated the modulating effects of astrocyte on neural cluster synchronization. It is found that a subset of neurons on the network can be synchronized into a cluster and, during the process of system evolution, the cluster is switching intermittently between the synchrony and asynchrony states at a slow frequency, namely the phenomenon of cluster breathing. By the method of symmetry-based analysis and by the technique of FLE, we have conducted a theoretical analysis on the stability of the cluster and found that the breathing activities are induced by the interplay between the astrocyte and the neural network. The generality of the breathing phenomenon has been confirmed in different network models, and the dependence of the breathing properties on the astrocyte parameters has been analyzed. The findings give insights into the regulating effects of astrocytes on neural firing activities, and are helpful to our understanding on the state switching of the cortical network.

This work was supported by the National Natural Science Foundation of China (NNSFC) under Grant Nos.~12275165, 11074159, and 11374199. X.G.W. was also supported by the Fundamental Research Funds for the Central Universities under Grant No. GK202202003.

\end{document}